\documentclass[11pt,a4paper]{article}
\usepackage{jheppub}
\usepackage[english]{babel}
\usepackage{graphicx}
\usepackage{longtable}

\newcommand{\mathsym}[1]{{}}

\title{Quark masses and mixings in the RS1 model with a condensing 4th generation}
\author{A. E. C\'arcamo Hern\'andez, }
\emailAdd{antonio.carcamo@usm.cl}
\author{Claudio O. Dib, }
\emailAdd{claudio.dib@usm.cl}
\author{Nicol\'as Neill H}
\emailAdd{nicolas.neill@gmail.com}
\author{and Alfonso R. Zerwekh}
\emailAdd{alfonso.zerwekh@usm.cl}
\affiliation{Universidad T\'ecnica Federico Santa Mar\'{\i}a\\
and\\
Centro Cient\'{\i}fico-Tecnol\'ogico de Valpara\'{\i}so\\
Casilla 110-V, Valpara\'{\i}so, Chile}
\date{\today }

\abstract{
We study the hierarchy of quark masses and mixings in a model based on a 5-dimensional spacetime with constant curvature of Randall-Sundrum type with two branes, where the Electroweak Symmetry Breaking is caused dynamically by the condensation of a 4th generation of quarks, due to underlying physics from the 5D bulk and the first KK gluons.
We first study the hierarchy of quark masses and mixings that can be obtained from purely adjusting the profile localizations, finding that realistic masses are not reproduced unless non trivial hierarchies of underlying 4-fermion interactions from the bulk are included. Then we study global $U(1)$ symmetries that can be imposed in order to obtain non-symmetric modified Fritzsch-like textures in the mass matrices that reproduce reasonably well quark masses and CKM mixings.
}

\begin{document}
\maketitle

\section{Introduction}\label{sec:intro}

One of the outstanding unresolved issues in Particle Physics is the origin of the masses of fundamental fermions.
The current theory of strong and electroweak interactions, the Standard Model (SM), has proven to be remarkably successful in passing all experimental tests \cite{PDG}. However, many people consider it to be just an effective framework of a yet unknown more fundamental theory, for several reasons. One of them is the hierarchy problem that arises from the quadratic divergence of the Higgs mass, indicating that there must be a so far unknown underlying physics in the gauge symmetry breaking mechanism. Another reason, related to the previous one, is the lack of an explanation for the large hierarchy of the fermion masses, which spread over a range of five orders of magnitude in the quark sector, and a dramatically broader range if we include the neutrinos. The origin of quark mixing and the size of CP violation in this sector is also a related issue. A fundamental theory, one expects, should have a dynamical explanation for the masses and mixings. Even though in the SM these parameters appear only through Yukawa interaction terms and not in explicit mass terms, this mechanism does not really provide an explanation for their values but only translates the problem to fitting different Yukawa couplings, one for each mass and with just as disparate values.

Several proposals to infer the pattern of fermion masses and mixings exist \cite{Fritzsch,Fx,Xing2011,Matsuda,Branco,Zhou,Carcamo,String,GUT,Extradim,Ncopies,Chivukula}, based on extended symmetries in the context of Two Higgs Doublets, Grand Unification, Extra Dimensions, Superstrings, Technicolor or $N$ copies of the SM, each of which generate specific textures for the fermion masses. One clear and outstanding feature in the pattern of quark masses is that they increase from one generation to the next, and that the mixings from the first to the second and to the third family are in decreasing order \cite{PDG,Fritzsch,Fx,Xing2011}.

Here we want to study the hierarchy of quark masses and mixings, using a specific framework of Electroweak Symmetry Breaking (EWSB) that has been recently proposed by Burdman and Da Rold (BDR) \cite{Burdman1}, which is  based on a 5-D spacetime with constant curvature of a Randall-Sundrum type \cite{rs} with two branes, where the hierarchy problem is naturally solved by exponential factors that appear in the reduction to a  4-dimensional theory, due to the 5-D space curvature. The BDR model does not include a fundamental Higgs, but breaks the symmetry dynamically due to the condensation of a 4th generation of quarks. The quark masses also appear as a consequence of condensation, and their values are modulated by the localization of the fermion zero mode profiles in the 5th dimension.

In this framework, it would be useful to know whether the values of fermion masses appear just due to the localization of the profiles without further reference to the underlying physics, or whether additional ingredients are required. Here we have explored this possibility and found that the constraints determined by the profiles alone are not enough to build the fermion mass hierarchy, but details of the underlying physics
in the bulk of the 5-D spacetime are necessary as well.

In this model of dynamical symmetry breaking, the requirement of condensation implies large effective couplings, which in turn results in large masses. To agree with current data including high precision tests, both quarks in the 4th generation doublet must be heavy (the $m_T-m_B$ splitting should not exceed $\sim 100$ GeV) \cite{Alok,kribs,Erler}. The scenario where only $T$ quark condenses results in an unacceptably large $m_T-m_B$ splitting, consequently both $T$ and $B$ quarks must condense. Within this framework, we study the hierarchy of quark masses and mixings for different effective bulk interaction coefficients, freely scanning over a range around their naive values. After finding some hierarchies among the effective coefficients and showing that their naive values do not reproduce acceptable masses, we study the inclusion global $U(1)$ symmetries in the effective operators that imply the textures in the quark matrices that give reasonable masses and mixings.

The content of this paper is organized as follows. In Section \ref{sec:model} we briefly describe the general features of this 5D scenario of dynamical EWSB and generation of quark masses. In Section \ref{sec:Tcond} we show the results of free numerical explorations,
first considering adjustment of profiles only, and then including some hierarchy coming from
underlying interactions in the bulk.
In Section \ref{sec:TBcond} we derive textures for the quark mass matrices from global $U(1)$ symmetries in the interaction operators. In Section \ref{sec:conclusions} we state our conclusions.

\section{The model}\label{sec:model}

We consider the RS1 model based on a warped extra dimension
compactified on an ${S^{1}}/{Z_{2}}$ orbifold, which corresponds to the
interval $\left[ 0,\pi R\right] $, and with a metric of anti-de Sitter (AdS) type given by \cite{rs}:
\begin{equation}
ds^{2}=e^{-2ky}\eta _{\mu \nu }dx^{\mu }dx^{\nu }+dy^{2}.
\end{equation}

Here $\eta _{\mu \nu }=diag\left( -1,1,1,1\right) $ is the four-dimensional
Minkowski metric and $k$ the AdS curvature. Since the Planck mass $M_P\sim 10^{19}$ GeV is the fundamental scale, a natural theory
should have $k\sim  M_P$. The TeV scale can be generated at the brane located at $y= \pi R$ if the compactification radius $R$ is such that $ke^{-\pi kR}=1$ TeV, which in turn means $kR \simeq 12$, also a rather natural number.
The electroweak gauge symmetry in the bulk has to be extended to
$SU\left( 2\right) _{L}\times SU\left( 2\right) _{R}\times U\left( 1\right)_{B-L}$ in order to avoid a too large violation of the custodial symmetry caused by $U(1)_Y$ KK modes \cite{adms,Tasi,higgsless,acdp,delocal}.
The model we study \cite{Burdman1} is based on the gauge symmetry given above, with four generations of fermions that propagate in the 5-dimensional bulk. We restrict our study to the quark sector. Boundary conditions at the $y=0$ and $y=\pi R$ branes determine a KK tower of modes, of which the zero modes are assumed to be the SM fields in 4-D. Since there is no chirality in 5 dimensions, the left- and right-handed fermions in 4-D correspond to zero modes of different fields in the bulk, conventionally called $\Psi_R$ and $\Psi_L$, which obey the conditions $\Psi _{R,L}=\pm \gamma _{5}\Psi _{R,L}$. The masses of the zero modes in 4-D appear when the electroweak symmetry is spontaneously broken, as in the SM. However in this model the Higgs field is not fundamental but only an effective low-energy field, and the electroweak symmetry is broken by the condensation of the 4th generation quarks.
Our goal is to see whether a realistic spectrum of quark masses and mixing parameters can be obtained just by adjustment of profiles in the 5th dimension, or further ingredients from underlying physics in the bulk are necessary.

Fermion fields in the bulk of 5D space, generically denoted as $\Psi(x,y)$, obey an action of the form:

\begin{equation}
S_5^{(\Psi)} = \int d^4 x \int^{\pi R}_{0} dy \sqrt{-g} \left(   i\overline{\Psi} \Gamma^M \nabla_M \Psi + i m_\Psi \overline{\Psi} \Psi      + \frac{C^{ijkl}}{M_5^3} \overline{\Psi}_L^i {\Psi}_R^j
\overline{\Psi}_R^k {\Psi}_L^l      \right),\label{SF}
\end{equation}
where capital indices run over the five coordinates, $\Gamma ^{M}=(e^{ky}\gamma ^{\mu }, \gamma^5) $ are the
Gamma matrices in the 5D curved spacetime and $\nabla_M$ is the covariant derivative that includes the interaction with the gauge fields.
The fermion masses in 5-D have natural values of the order of the Planck mass:
\begin{equation}
m_{\Psi}
=k\ d_{\Psi} , \label{localizations}
\end{equation}
where $k$ is the AdS curvature and $d_{\Psi}$ are parameters of order unity. These parameters will determine the localization of the fermion profiles along the 5th dimension. The masses of fermions in the TeV brane (the observable 4D space) will be a consequence of this localization after
electroweak symmetry breakdown.  $M_{5}$ is the 5-D Planck mass, which is related to $M_P$, the Planck mass in 4-D, through the following relation:
\begin{equation}
M_{P}^{2}=\frac{M_{5}^{3}}{k}\left( 1-e^{-2\pi kR}\right).
\end{equation}
Since the exponential term is highly suppressed, $M_5$,  $M_P$ and $k$ are all of the same order of magnitude.
%
%
%
%
%
%
%

The last term in Eq.\ (\ref{SF}) is an effective 4-fermion term that is assumed to arise from underlying interactions in the 5-D bulk below the Planck scale; the effective coefficients $C^{ijkl}$ run over all flavors in a way that respect the electroweak gauge symmetry, and have values that can be estimated by naive dimensional analysis (NDA).

The NDA estimate of the coefficients in a low energy effective theory consists, in general, in requiring that the 4-fermion effective operator should have a coefficient at tree level,  $C^{ijkl}/\Lambda^3$, of the same magnitude as its 1-loop contribution at the cutoff scale $\Lambda$ (in this case, $\Lambda = M_5$). Since the 1-loop contribution has two vertices (in this case), it is quadratic in the effective coefficient, while the tree level term is linear, and consequently the relation for the coefficient is non-trivial. The naive estimate of the loop integral includes the mass dimensions as well as the factors of $2\pi$:
\[
\frac{C} {\Lambda^3}    
  \sim  \left( \frac{C}{\Lambda^3} \right)^2  N \int^\Lambda  \frac{d^5 k}{(2\pi)^5} \left( \frac{1}{k}\right)^2 ,
\]
with $N\sim 80$ being the number of fermions in the theory that run inside the loop. 
Doing the cutoff integral and solving for $C$ one obtains $C\sim 36\pi^3/N$.


Fermion and gauge boson fields can be expanded in their respective KK modes as follows:
\begin{eqnarray}
\Psi _{L,R}\left( x,y\right) &=&\sum_{n=0}^{\infty }\Psi
_{L,R}^{\left( n\right) }\left( x\right) f_{L,R}^{\left( n\right) }\left(
y\right) ,\nonumber\\
A_{\mu }^{a}\left( x,y\right) &=&\sum_{n=0}^{\infty }A_{\mu
}^{\left( n\right) }\left( x\right) \chi^{\left( n\right) }\left( y\right) ,
\end{eqnarray}
where their profiles in the 5th dimension are respectively given by \cite{Tasi}:
\begin{equation}
f_{L,R}^{\left( n\right) }\left( y\right) =
\left\{ \begin{array}{ll}
\sqrt{\frac{k\left(
1-2d_{L,R}\right) }{e^{\left( 1-2d_{L,R}\right) k\pi R}-1}}e^{\left(
2-d_{L,R}\right) ky}    & , \ n=0\\
N_{\Psi _{L,R}}^{\left( n\right)
}e^{\frac{5}{2}ky} \left[
J_{\pm d_{L,R}\pm \frac{1}{2}}
\left(
\frac{  m_{\Psi}^{\left( n\right) } }  {k} e^{ky}
\right)
+
\alpha _{\Psi _{L,R}}^{\left( n\right) }
Y_{\pm d_{L,R}\pm \frac{1}{2}}
\left(
\frac{ m_{\Psi}^{\left( n\right) } }  {k} e^{ky}
\right) \right]  & , \ n>0 ,
\end{array}\right.
\end{equation}
\begin{equation}
\chi^{\left( n\right) }\left(y\right)
= \left\{ \begin{array}{ll}
\frac{1}{\sqrt{\pi R}} & , \ n=0\\
N_{A}^{\left( n\right) }e^{ky}\left[
J_{1}
\left(
\frac{ m_A^{\left( n\right) } }  {k} e^{ky}
\right)
+
\alpha_{A}^{\left( n\right) }
Y_{1}
\left(
\frac{m_A^{\left( n\right) } } {k} e^{ky}%
\right) \right] & , \ n>0 .
\end{array}\right.
\end{equation}
%
Here $\chi^{\left( n\right) }\left(y\right)$ are given for the gauge $A_5 =0$. 
%
The functions $J_{\rho}$, $Y_{\rho}$ are first and second kind Bessel functions, respectively and $N_{\Psi _{L,R}}^{\left( n\right) }$, $N_{A}^{\left( n\right) }$ are normalization constants computed from the following orthonormality relations:
\begin{equation}
\int^{\pi R}_{0} dy e^{-3ky} f_{L,R}^{\left( n\right) }\left(y\right)f_{L,R}^{\left( m\right) }\left(y\right)=\delta_{nm},\hspace{1.5cm}\int^{\pi R}_{0} dy\chi^{\left( n\right) }\left( y\right)\chi^{\left( m\right) }\left( y\right)=\delta_{nm}
\end{equation}
and the coefficients $\alpha _{\Psi _{L,R}}^{\left( n\right) }$ and $\alpha_{A}^{\left( n\right) }$ are determined by the boundary conditions on the branes, resulting in the following relations:
\begin{eqnarray}
\alpha _{\Psi _{L,R}}^{\left( n\right) }&=&
-\frac{J_{\pm d_{L,R}\pm \frac{1}{2}}
\left( \frac{m_\Psi ^{\left( n\right) }}{k}\right) }
{Y_{\pm d_{L,R}\pm
\frac{1}{2}}\left( \frac{m_\Psi ^{\left( n\right) }}{k}\right) }
=
-\frac{J_{\pm d_{L,R}\pm \frac{1}{2}}
\left( \frac{m_\Psi ^{\left( n\right) }}
{k}e^{\pi kR}\right) }
{Y_{\pm d_{L,R}\pm \frac{1}{2}}
\left( \frac{m_\Psi ^{\left( n\right) }}{k}e^{\pi kR}\right) },\\
\alpha _{A}^{\left( n\right) }&=&
-\frac{J_{0}\left( \frac{m_A^{\left( n\right) }}{k}\right) }
{Y_{0}\left( \frac{m_A^{\left( n\right) }}{k}%
\right) }
\quad =
-\frac{J_{0}\left( \frac{m_A^{\left( n\right) }}{k}e^{\pi kR} \right) }
{Y_{0}\left( \frac{m_A^{\left( n\right) }}{k}e^{\pi kR}\right) }.\label{alphaA}
\end{eqnarray}
In turn, these relations determine the masses $m_\Psi ^{\left( n\right) }$ and $m_A^{\left(
n\right) }$ of the non zero modes. These modes correspond to heavy particles beyond the spectrum of the Standard Model (SM). On the other hand, the zero modes, which are identified with the SM fermions remain massless at this level and become massive only after EWSB.

In the model under consideration, EWSB is caused dynamically by condensation of the 4th generation $T$ and $B$ quarks.
The composite scalar sector generated by this condensation pattern corresponds
to that of a Two Higgs Doublet Model (2HDM). Fourth generation leptons have to be added as well in order to avoid gauge anomalies \cite{Burdman2}, but since they are color singlets, they are assumed not to experience condensation.


The dynamical masses of the 4th generation $T$ and $B$ quarks are related to the symmetry breaking
scale $v=246$ GeV through the following Pagels-Stokar relation:
\begin{equation}
v^{2}=\frac{N_{C}}{8\pi ^{2}}\left[ m_{T}^{2}\ln \left( \frac{M_{KK}^{2}}{%
m_{T}^{2}}\right) +m_{B}^{2}\ln \left( \frac{M_{KK}^{2}}{m_{B}^{2}}\right) %
\right].\label{PS2}
\end{equation}
where $N_C=3$ is the number of colors and $M_{KK}$ is the mass of the first Kaluza-Klein gluon. Here we use $M_{KK}=2.4$~TeV, which is obtained from Eq. (\ref{alphaA}) requiring that $kR \sim 12$. The masses of the 4th generation $T$ and $B$ quarks are constrained by
electroweak precision tests. The minimal contribution to the oblique
parameters $\widehat{S}$ and $\widehat{T}$ leads to a splitting $m_{T}-m_{B} \lesssim 100$ GeV \cite{Alok,kribs,Erler}. Since the size of quark masses are related to the strength of the
interaction that causes the condensation, the scenario where only $T$ quark condenses results in $m_T-m_B$ splittings much larger than the value cited above. Consequently both $T$ and $B$ quarks must condense. In our numerical exploration we will fix $m_{T}-m_{B} = 55$ GeV, as in Ref. \cite{Alok}. By combining this choice with the Pagels-Stokar relation in Eq. (\ref{PS2}), it follows that our 4th generation quark masses are $m_{T}=528$ GeV and $m_{B}=473$ GeV.

The interaction responsible for the condensation of the $T$ and $B$ quarks is assumed to be a combination of an effective 4-fermion interaction arising from the bulk [last term in Eq.\ (\ref{SF})] and a similar interaction arising from the exchange of the first KK gluons.

Let us now describe the condensation mechanism. The full effective interaction that causes the $T$ and $B$ condensation is obtained after integrating the effective 4-fermion interaction terms in the action over the 5th coordinate, $y$. These effective terms arise from underlying interactions in the bulk of 5-D spacetime as well as from the exchange of the first KK gluons (we assume heavier KK gluon terms are comparatively smaller). The integration over $y$ generates effective couplings in 4-D that depend on the profile localization parameters $d_\Psi$ [see Eq.\ (\ref{localizations})]. The resulting expression for the 4-$T$ and 4-$B$ interaction is:
\begin{eqnarray}\label{MKK}
{\cal L}_{\psi^{4}}&=&\frac{C^{TTTT}}{M_{P}^{2}}f_{TT}^{\left( T\right) }%
\left[\overline{T}_{L}(x) T_{R}(x)\right] \left[\overline{T}_{R}(x) T_{L}(x)\right]\nonumber\\
&&+\frac{C^{BBBB}}{M_{P}^{2}}f_{BB}^{\left( B\right) }%
\left[\overline{B}_{L}(x) B_{R}(x)\right] \left[\overline{B}_{R}(x) B_{L}(x)\right]\nonumber
\\
&& -\frac{g_{T}^{L}g_{T}^{R}}{M_{KK}^{2}}\left[ \overline{T}_{L}(x) \gamma _{\mu }\frac{\lambda ^{a}}{2}T_{L}(x) \right]
\left[ \overline{T}_{R}(x) \gamma ^{\mu }\frac{\lambda ^{a}}{2}T_{R}(x) \right]\nonumber
\\
&&-\frac{g_{B}^{L}g_{B}^{R}}{M_{KK}^{2}}\left[ \overline{B}_{L}(x) \gamma _{\mu }\frac{\lambda ^{a}}{2}B_{L}(x) \right]
\left[ \overline{B}_{R}(x) \gamma ^{\mu }\frac{\lambda ^{a}}{2}B_{R}(x) \right]+\ldots,
\end{eqnarray}
where the ellipsis refer to other 4-quark terms. Here $\lambda ^{a}$ are the Gell Mann matrices,
the coefficients $f_{QQ}^{\left( Q\right) }$ ($Q=T,B$) are obtained after the $y$-integration over the profiles:
\begin{equation}
f_{QQ}^{\left( Q\right) }=\frac{\left( 1-2d_{Q_{L}}\right) \left(
1-2d_{Q_{R}}\right) \left( 1-e^{-2\pi kR}\right) \left[ e^{\left(
4-2d_{Q_{L}}-2d_{Q_{R}}\right) k\pi R}-1\right] }{2\left(
2-d_{Q_{L}}-d_{Q_{R}}\right)\left[ e^{\left(
1-2d_{Q_{L}}\right) k\pi R}-1\right] \left[ e^{\left(
1-2d_{Q_{R}}\right) k\pi R}-1\right]},
\end{equation}
and, similarly,  $g_{T}^{L,R}$ and $g_{B}^{L,R}$ are the left-/right-handed effective couplings of the $T$ and $B$ quarks to the first KK gluon:
\begin{equation}
g_{Q}^{L,R}=g_{s}\frac{\left( 1-2d_{Q_{L,R}}\right) k\sqrt{\pi R}}{e^{\left(
1-2d_{Q_{L}}\right) k\pi R}-1}\int_{0}^{\pi R}dy\  e^{\left(
1-2d_{Q_{L,R}}\right) ky}  \  \chi^{\left( 1\right) }\left( y\right), \ \ \ \ \ \ \ Q=T,B,\label{kkgluon}
\end{equation}
where $g_s$ is the QCD coupling. 
By performing a Fierz rearrangement in Eq. (\ref{MKK}), the effective four-fermion interaction Lagrangian responsible for the condensation of $T$ and $B$ quarks can be rewritten as follows:
\begin{equation}
{\cal L} _{\psi^{4}}= \frac{g_{T}^{2}}{M_{KK}^{2}} \left[\overline{T}_{L}T_{R}\right] \left[\overline{T}_{R}T_{L}\right]\ +\frac{g_{B}^{2}}{M_{KK}^{2}} \left[\overline{B}_{L}B_{R}\right] \left[\overline{B}_{R}B_{L}\right] \  +      {\cal O}(1/N_c) +\ldots .\label{effL}
\end{equation}
Here we have omitted crossed terms of the form $\left[\overline{T}_{L}T_{R}\right]\left[\overline{B}_{R}B_{L}\right]+ h.c.$, as they will be forbidden by the $U(1)$ symmetries we will impose (see section \ref{sec:TBcond}). Besides that, we have defined the effective coupling $g_{Q}^{2}$ as:
\begin{equation}
g_{Q}^{2}=g_{Q}^{L}g_{Q}^{R}+C^{QQQQ}f_{QQ}^{\left( Q\right) }\frac{M_{KK}^{2}}{M_{P}^{2}},\ \ \ \ \ \ \ Q=T,B.
\end{equation}
Considering Eq.\ (\ref{effL}) as an effective interaction below the scale $M_{KK}$, the condensates $\left\langle 0|\overline{T}_{L}T_{R}|0\right\rangle\ne 0$ and $\left\langle 0|\overline{B}_{L}B_{R}|0\right\rangle\ne 0$ form when the couplings $g_T$ and $g_B$, respectively, satisfy the condition \cite{bhl}:
\begin{equation}
g^{2}_T,g^{2}_B>\frac{8\pi^{2}}{N_C}.\label{cb}
\end{equation}
%
%
%

%
 These couplings become strong due to the presence of bulk interactions, formulated in terms of the effective couplings
$C^{ijkl}$ in Eq.\ (\ref{SF}), and to the strength of the fermion couplings to the first KK gluon shown in Eq.\ (\ref{kkgluon}),
both of which depend on
the localization of the profiles in the fifth dimension, $d_{Q_L}$ and $d_{Q_R}$ ($Q=T,B$).
The appearance of these condensates spontaneously breaks the gauge symmetry, and generates dynamical masses for the quarks that condense. In this case, two composite Higgses appear with non vanishing vacuum expectation values, resulting in the breaking of the Electroweak Symmetry and in the subsequent generation of quark masses. Specifically, the dynamical masses of the condensing quarks are related to the condensates:
\begin{equation}
m_{T}=-g_T^2 \frac{\langle \overline T_L T_R \rangle}{M_{KK}^2}, \qquad m_{B}=-g_B^2 \frac{\langle \overline B_L B_R \rangle}{M_{KK}^2}.
\end{equation}
Using $M_{KK} = 2.4$ TeV (see above) and taking Eq.\ (\ref{cb}) into account, a good numerical approximation for these relations is $\left\langle \overline{T}_{L}T_{R}\right\rangle=-\kappa m^{3}_T$ and $\left\langle \overline{B}_{L}B_{R}\right\rangle=-\kappa m^{3}_B$, where $\kappa$ is a number of order $O(1)$ \cite{Burdman1}. Here we use $\kappa\simeq 0.8$, which is consistent with both condensation conditions in  Eq. (\ref{cb}). Now, concerning the masses of the first three generations of quarks, we turn to the effective 4-fermion operators that contain just a $T\overline T$ or a $B\overline B$ pair:
%
%
%
\begin{equation}
{\cal L}_{\Psi ^{4}} =C^{ijTT} \frac{f_{ij}^{\left( T\right) }}{M_{P}^{2}}\
\left[\overline{T}_{L}T_{R}\right] \left[\bar{q}_{iR}q_{jL}\right]
+ C^{ijBB} \frac{f_{ij}^{\left( B\right) }}{M_{P}^{2}}\
\left[\overline{B}_{L}B_{R} \right]\left[ \bar{q}_{iR}q_{jL} \right]+ \ldots
\end{equation}
where the coefficients $f^{(Q)}_{ij}$ ($Q=T,B$) arise from the integrals of the profiles over $y$, and are given by
\begin{eqnarray}
f_{ij}^{(Q)} &=&
\frac{\exp\left\{\left(
4-d_{i_L}-d_{j_R}-d_{Q_{L}}-d_{Q_{R}}\right)
k\pi R\right\}-1}{4-d_{i_{L}}-d_{j_{R}}-d_{Q_{L}}-d_{Q_{R}}}%
\\
&&\times
\sqrt{\frac{\left( 1-2d_{i_{L}}\right) \left(
1-2d_{j_{R}}\right) }{\left[ \exp\left\{\left( 1-2d_{i_{L}}\right)
k\pi R\right\}-1\right] \left[ \exp\left\{\left( 1-2d_{j_{R}}\right) k\pi R\right\}-1\right] }}\notag \\
&&\times \sqrt{\frac{\left( 1-2d_{Q_{L}}\right) \left(
1-2d_{Q_{R}}\right) }{\left[ \exp\left\{\left( 1-2d_{Q_{L}}\right) k\pi R\right\}-1%
\right] \left[ \exp\left\{\left( 1-2d_{Q_{R}}\right) k\pi R\right\}-1\right] }}
\notag \\
&&\times
\left(1-e^{-2\pi kR}\right).\label{e:fij}
\end{eqnarray}
Here the indices $i,j$ run over the quark flavors, either $u,c,t$ for the up sector or $d,s,b$ for the down sector.

After condensation, these interactions generate the quark mass matrices for the first 3 generations:
\begin{equation}
M_{ij} =  C^{ijTT}\frac{f_{ij}^{(T)}}{M_P^2} \kappa \ m_T^3  +
C^{ijBB}\frac{f_{ij}^{(B)}}{M_P^2} \kappa \ m_B^3  . \label{emumd}
\end{equation}
%
On the other hand, the coefficients $C^{ijkl}$, introduced already in Eq. (\ref{SF}), are effective parameters related to underlying interactions in the bulk of 5-D; in naive dimensional analysis, these coefficients have a value $\sim 36\pi^3 / N$, where $N$ is the number of fermions in the theory.

In the next sections we do numerical analyses to study the conditions under which these quark mass matrices can reproduce the measured values of quark masses and mixings. First we do a general scanning of the $C^{ijkl}$ parameters that reproduce the physical masses and then we look for $U(1)$ symmetries that impose specific textures on the mass matrices.

\section{Quark mass hierarchy and non-universality of 4-fermion interactions}\label{sec:Tcond}

After condensation of the 4th generation of quarks, the mass matrices for the first three generations take the form given in Eq.\ (\ref{emumd}).
We will show that if all the coefficients $C^{ijTT}$ and $C^{ijBB}$ are equal, as assumed in a naive dimensional analysis, these
mass matrices cannot reproduce realistic values for the quark masses. The generation mixing implied in Eq.\ (\ref{emumd})
makes these matrices too singular. Consequently, by fitting the heavy eigenvalues to the 3rd generation, the prediction for
the first two generations turns out to be unacceptably small.

The mass matrices $M^{u}$ and $M^{d}$ (up and down sector, respectively) can be diagonalized by a bi-unitary transformation
\begin{eqnarray}
M^{q}_{D}=V^{q}_L M^{q} V^{q\dag}_R,\ \ q= u,d ,
\end{eqnarray}
where $V^{q}_{L,R}$ are unitary matrices.  We know that the determinant of
the mass matrices satisfies $|\mbox{det} M^{q}| = |\mbox{det} M^{q}_D|$.
Let us take $M^d$ as an example. For this matrix
\begin{eqnarray} \label{e:detmu}
\det M^d_D = m_d m_s m_b \approx 4.6\times 10^{-4}\mbox{ GeV}^3,
\end{eqnarray}
thus any $M^d$ matrix that correctly reproduces the quark masses must satisfy this approximate condition.

Now we proceed to do a numerical study of the space of localization parameters, $d_{q_L}$ and $d_{q_R}$, in order to find realistic solutions for the quark masses, keeping the bulk coefficients
$C^{ijQQ}$ ($Q=T,B$) equal to their naive value. We first search for the parameter space which correctly fits the value of the bottom quark mass, taken to be $m_b\approx 2.9\mbox{ GeV}$.
In order to reduce the parameter space in the calculation, we restrict our search to
two limiting scenarios \cite{Gherghetta:2000qt}.
Scenario 1: left-right symmetry in the localization parameters, i.e. $d_{i_L}=d_{i_R}$ ($i=d,s,b$) and
Scenario 2: all left handed localization parameters are fixed at the point $d_{i_L}=1/2$ ($i=d,s,b$); this is the point at which the profiles are flat, i.e. not localized towards either brane.
In both of these scenarios the bottom quark mass is computed by varying the right-handed localization profiles in the following ranges
\begin{equation}
0 \leq d_{b_R} \leq 0.9 , \ \ \  0.5 \leq d_{d_R},d_{s_R} \leq 1.5 . \nonumber
\end{equation}
For the parameter space described above, good fits for the bottom quark mass are obtained when its right-handed profile parameter lies in the ranges
\begin{eqnarray} \label{e:cotadt}
\begin{array}{rcl}
0.54 < & d_{b_R},&\mbox{  (Scenario 1)}, \\
0.57 < & d_{b_R},&\mbox{  (Scenario 2)} .
\end{array}
\end{eqnarray}

Since the mass matrices in Eq.\ (\ref{emumd}) approach a singular structure when all $C^{ijQQ}$ are equal, ($= 36\pi^3/N$, see previous Section), we look for matrices $M^d$ that successfully account for the correct value of the bottom quark mass (i.e. we limit $d_{b_R}$  to the ranges given in Eq. \ref{e:cotadt}), and at the same time give maximal values for the determinant [see Eq. (\ref{e:detmu})].
We found that in both Scenarios the maximum of the determinant is a decreasing function of $d_{b_R}$, consequently the largest determinant values are found at the lowest $d_{b_R}$ values, namely  $d_{b_R}=0.54$ and $d_{b_R}=0.57$, for the aforementioned
scenarios, respectively.
The corresponding determinants at these points are:
\begin{eqnarray}
1.2\times 10^{-9}\mbox{ GeV}^3,&\mbox{  (Scenario 1)},\nonumber\\
2.9\times 10^{-11}\mbox{ GeV}^3,&\mbox{  (Scenario 2)}.\nonumber
\end{eqnarray}

These results are $5$ and $7$ orders of magnitude below the required value $\det M^d\approx 4.6\times 10^{-4}\mbox{ GeV}^3$, respectively.
This means, for example, that if we would correctly reproduce the strange and bottom quark masses, at best we will
get a down quark mass $5$ orders of magnitude below its real value, $m_d\approx 1-5\mbox{ MeV}$ (at the $M_Z$ scale). We therefore conclude that it is not possible to reproduce a realistic spectrum of quark masses by adjusting the profiles only, keeping all $C^{ijQQ}$ coefficients equal. We should recall that these are effective coefficients coming from the underlying physics in the 5-D bulk.

We now proceed to study whether there are non-trivial hierarchies in the $C^{ijQQ}$ coefficients
that, together with the adjustment of profiles, can lead to realistic predictions for quark masses and mixings.

We try the following hierarchy parametrization for the  $C^{ijQQ}$ coefficients:

\begin{eqnarray}
C^{ijQQ} \approx 36\pi^3/N\times
\left\{
\begin{array}{l}
1,\ i=j,\\
C^{ij},\ i\neq j,
\end{array}\right.
\end{eqnarray}
%
where $i,j=1,2,3$ and $C^{ij}$ are expected to be of order $O(1)$ or possibly lower.

\begin{table}[tbh]
\begin{center}
\begin{tabular}{c|c|c|c}
\hline\hline
Quark flavour & Mass $m_q$ & Left profile $d_{q_L}$ & Right profile $d_{q_R}$  \\ \hline
u & $1.27$  MeV & $0.605$  & $0.712$  \\ \hline
d & $2.9$   MeV & $0.605$  & $0.629$  \\ \hline
c & $619$   MeV & $0.514$  & $0.405$  \\ \hline
s & $55$    MeV & $0.514$  & $0.689$  \\ \hline
t & $172$   GeV & $-1.117$ & $0.631$  \\ \hline
b & $2.89$  GeV & $-1.117$ & $0.689$  \\ \hline
T & $528$   GeV & $0.0783$  & $0.209$  \\ \hline
B & $473$   GeV & $0.0783$  & $0.109$  \\ \hline\hline
\end{tabular}%
\end{center}
\caption{Quark profile values that reproduce the given quark masses and the best fit for the
magnitudes of the CKM mixing elements.
The corresponding $C^{ij}$ coefficients obtained in this fit are $C^{12}=0.008$, $C^{13}=0.486$, $C^{21}=0.217$, $C^{23}=0.596$, $C^{31}=0.731$ and $C^{23}=0.451$.}
\label{t:locs1}
\end{table}

\begin{table}[bh]
\begin{center}
\begin{tabular}{c|c|c}
\hline\hline
CKM matrix element & Obtained Value & Experimental Value \\ \hline
$\bigl|V_{ud}\bigr|$ & $0.974$ & $0.97426\pm 0.00030$ \\ \hline
$\bigl|V_{us}\bigr|$ & $0.225$ & $0.22545\pm 0.00095$ \\ \hline
$\bigl|V_{ub}\bigr|$ & $0.00402$ & $0.00356_{-0.00020}^{+0.00015}$ \\ \hline
$\bigl|V_{cd}\bigr|$ & $0.225$ & $0.22529\pm 0.00077$ \\ \hline
$\bigl|V_{cs}\bigr|$ & $0.973$ & $0.973416_{-0.00018}^{+0.00021}$ \\ \hline
$\bigl|V_{cb}\bigr|$ & $0.0486$ & $0.0493_{-0.00528}^{+0.00075}$ \\
\hline
$\bigl|V_{td}\bigr|$ & $0.00700$ & $0.00861_{-0.00037}^{+0.00021}$ \\ \hline
$\bigl|V_{ts}\bigr|$ & $0.0482$ & $0.040681_{-0.00138}^{+0.00043}$ \\ \hline
$\bigl|V_{tb}\bigr|$ & $0.999$ & $0.999135_{-0.000018}^{+0.000057}$ \\
\hline\hline
\end{tabular}%
\end{center}
\caption{Obtained and experimental values of the magnitudes of the CKM
matrix elements.}
\label{t:ckm1}
\end{table}

In this new study we consider the quark mass matrices for both the ``up'' and ``down'' sectors. We want to find the sets of values for the above $C^{ij}$ coefficients that correctly reproduce the quark mass spectrum for some localization of the profiles. We scan over the $C^{ij}$ coefficients, looking for those that give correct
eigenvalues for the quark mass matrices.
However, it is not enough to fix $C^{ij}$ values to get unique mass eigenvalues: for the first three generations we have $9$ localization parameters, $6$ corresponding to the right handed quarks and $3$ to the left handed quarks. Notice that the $SU(2)_L$ symmetry restricts the left handed localizations of ``up" and ``down" type quarks of a given family to be equal, i.e. $d_{u_L}=d_{d_L}$, $d_{c_L}=d_{s_L}$, $d_{t_L}=d_{b_L}$.
Since we need to fit 6 masses, there are still 3 localization parameters to be chosen at will, which we take to be $d_{u_L}$, $d_{c_L}$ and $d_{t_L} $.
Then, our procedure goes as follows: we scan over the $C^{ij}$ and the $d_{q_L}$ and, at each step, we check for the existence of a solution of the mass eigenvalue equation that reproduces the correct quark masses. We find that there are $C^{ij}$ values that do lead to correct solutions for the masses, while other values do not.
The parameters $C_{ij}$ and $d_{q_L}$ are scanned over the following ranges:
\begin{eqnarray}
0 \leq & C^{ij} & \leq 1, \nonumber\\
0.5 \leq & d_{u_L} & \leq 0.8,  \\
0.5 \leq & d_{c_L} & \leq 0.8,\nonumber\\
-2 \leq & d_{t_L} & \leq 0.5.\nonumber
\end{eqnarray}
%

The quark masses we use as reference in our fits, taken at the $M_Z$ scale \cite{Xing}, are shown in Table \ref{t:locs1}.

We found that approximately 15\% of the scanned range correctly reproduces all quark masses. Among these solutions, the one shown in Table \ref{t:locs1} produces the best agreement with the experimental values of the magnitudes of the Cabibbo-Kobayashi-Maskawa (CKM) matrix. Table \ref{t:ckm1} shows the magnitudes of the CKM elements we obtain at the parameter point just indicated.

This solution exhibits just a moderate hierarchy among the values of the $C^{ij}$ coefficients: $C^{12}=0.008$, $C^{13}=0.486$, $C^{21}=0.217$, $C^{23}=0.596$, $C^{31}=0.731$ and $C^{23}=0.451$. The main conclusion of this analysis is that some level of detail of the $C^{ijQQ}$ coefficients coming from the underlying physics in the 5D bulk is necessary to reproduce the hierarchy of the known quark masses.

\section{Quark masses and mixings from textures in the bulk}\label{sec:TBcond}

As we have seen in the previous section, the BDR model based on a condensing 4th generation of quarks does not provide a successful prediction for the SM quark masses when all  $C^{ijQQ}$  coefficients are assumed to be equal.  This assumption on the  $C^{ijQQ}$ leads to extremely low values for the first and second generation quark masses, if one adjusts the parameters to fit the top and bottom masses. We also found that this problem is solved when a hierarchy among the $C^{ijQQ}$ is introduced. This hierarchy must originate from the underlying physics in the 5-D bulk and, although rather mild, it seems to be required, in addition to the localization of the fermion profiles,
in order to explain the observed quark mass and mixing pattern. So far we do not have an explanation for the required hierarchy.

In this section we want to explore possible symmetries that could impose \emph{textures} in the $C^{ijQQ}$ coefficients. In what follows, we assume each of the $C^{ijkl}$ to equal $C$, or zero if there is a symmetry that forbids it. Now let us try to impose such symmetries.

We require that the 4-fermion interaction Lagrangian be invariant under the following $U(1)$ transformations:
%
\begin{equation}
\left(
\begin{array}{c}
u_{i} \\
d_{i}%
\end{array}%
\right) _{L}\rightarrow e^{i\alpha _{i}}\left(
\begin{array}{c}
u_{i} \\
d_{i}%
\end{array}%
\right) _{L},\hspace{0.5cm}\left( u_{i}\right) _{R}\rightarrow e^{i\beta
_{i}}\left( u_{i}\right) _{R},\hspace{0.5cm}\left( d_{i}\right)
_{R}\rightarrow e^{i\beta _{i}}\left( d_{i}\right) _{R},\hspace{0.5cm}i=1,2,3
\label{t}
\end{equation}

\begin{equation}
\left(
\begin{array}{c}
T \\
B%
\end{array}%
\right) _{L}\rightarrow e^{i\alpha _{4}}\left(
\begin{array}{c}
T \\
B%
\end{array}%
\right) _{L},\hspace{0.5cm}T_{R}\rightarrow e^{i\beta _{4}}T_{R},\hspace{%
1.5cm}B_{R}\rightarrow e^{i\lambda _{4}}B_{R},\hspace{0.5cm}\beta _{4}\neq
\lambda _{4},
\end{equation}
where all the phases $\alpha _{i}$ and $\beta _{i}$ ($i=1,2,3$) are different in general, but obeying the following restrictions:
\begin{equation}
\alpha _{4}-\beta _{4}=\alpha _{1}-\beta _{1}=\alpha _{2}-\beta _{3}=\alpha
_{3}-\beta _{2},\hspace{1.0cm}\alpha _{4}-\lambda _{4}=\alpha
_{1}-\beta _{2}=\alpha _{2}-\beta _{1}=\alpha _{3}-\beta _{3}.
\end{equation}
Now we can see that the condition $\beta_4 \neq \lambda_4$ forbids the mixing terms mentioned below Eq.\ (\ref{effL}).

Since the 4-fermion interaction Lagrangian should be invariant under the aforementioned $U(1)$ transformations, the following non-symmetric modified Fritzsch-like textures for the ``up" and ``down" type SM quarks are obtained:
\begin{eqnarray}
M^{u}&=&\frac{C  \kappa }{M_P^{2}}\left(\begin{array}{ccc}
f_{uu}^{\left( T\right) }m_{T}^{3} & f_{uc}^{\left( B\right) }m_{B}^{3} & 0 \\
f_{cu}^{\left( B\right) }m_{B}^{3} & 0 & f_{ct}^{\left( T\right) }m_{T}^{3} \\
0 & f_{tc}^{\left( T\right) }m_{T}^{3} & f_{tt}^{\left( B\right) }m_{B}^{3}\end{array}\right) ,
\nonumber
\\
& &\nonumber
\\
M^{d}&=&\frac{C \kappa}{M_{P}^{2}}%
\left(
\begin{array}{ccc}
f_{dd}^{\left( T\right) }m_{T}^{3} & f_{ds}^{\left( B\right) }m_{B}^{3} & 0 \\
f_{sd}^{\left( B\right) }m_{B}^{3} & 0 & f_{sb}^{\left( T\right) }m_{T}^{3} \\
0 & f_{bb}^{\left( T\right) }m_{T}^{3} & f_{bb}^{\left( B\right) }m_{B}^{3}
\end{array}%
\right).
\end{eqnarray}

Certainly at present there is no fundamental justification for these $U(1)$ symmetries, so they can be seen just as translations of the problem of textures. However they provide a better frame to organize
the textures and discover patterns and corrections in the mass and mixing matrices.

Now we proceed to scan over the profile parameters of the first 3 generation quarks, looking for the points that best reproduce their physical masses. As before, we use the quark masses at the $M_{Z}$ scale as reference values \cite{Xing}, which are shown in Table \ref{table1}.
Recalling that the $SU(2)_L$ symmetry restricts the left handed profile parameters of up- and down-type quarks of a given family to be equal ($d_{u_L} = d_{d_L}$, etc.), and for parameters $d>1/2$ or $d<1/2$ the localization goes towards the Planck brane or the TeV brane, respectively, we scan these parameters over the range $-1 \leq d_{q_L}, d_{q_R} \leq 1$,
except for the left-handed 3rd generation, where we use $-1.5 \leq d_{t_L}\leq 1$.
%

We scanned over this parameter space, trying to find points where the mass eigenvalues as well as the magnitudes of the CKM mixing elements fit the respective experimental values. The fitted profile parameters as well as the quark masses are shown in Table \ref{table1}, while the fitted CKM magnitudes are shown in Table \ref{table2}. As a consistency check, we have also verified that, for the quark profile parameters given in Table \ref{table1}, the third generation quarks do not condense. The obtained quark masses are in excellent agreement with their corresponding reference values at the $M_{Z}$ scale with the exception of the first generation ($u$ and $d$ quarks) which turn out to be larger by a factor $\sim 5$, an acceptable error given the large theoretical uncertainties involved in the light quark masses. In spite of this discrepancy, this result represents an improvement over the original scenario, where all $C^{ijkl}$ are taken at their naive value, which predicts first generation quark masses several orders of magnitude too small. This result coincides with the conclusion of the previous section, in the sense that details of the underlying physics in 5-D are crucial to reproduce the observed low energy phenomenology.

\begin{table}[tbh]
\begin{center}
\begin{tabular}{c|c|c|c|c}
\hline\hline
Quark flavour & Left profile $d_{q_L}$ & Right profile $d_{q_R}$ & Obtained
mass $m_q$ & Reference mass $m_q$ \\ \hline
u & $0.479$ & $0.265$ & $5.4$ MeV & $1.27_{-0.42}^{+0.50}$ MeV \\
\hline
d & $0.479$ & $0.602$ & $12.4$ MeV & $2.9 \pm 1.2$ MeV
\\ \hline
c & $0.025$ & $0.665$ & $626$ MeV & $619\pm 84$  MeV \\ \hline
s & $0.025$ & $0.745$ & $55.1$ MeV& $55 \pm 15$ MeV \\ \hline
t & $-1.43$ & $0.772$ & $181$ GeV & $172 \pm 3$ GeV \\ \hline
b & $-1.43$ & $0.739$ & $3.05$ GeV & $2.89\pm 0.09$ GeV \\ \hline
T & $0.0783$ & $0.209$ & $528$ GeV &  \\ \hline
B & $0.0783$ & $0.109$ & $473$ GeV &  \\ \hline\hline
\end{tabular}%
\end{center}
\caption{Quark profile values with the obtained and reference values of
quark masses at the $M_{Z}$ scale.}
\label{table1}
\end{table}
Table \ref{table2} shows the magnitudes of the CKM matrix elements we obtain in our parameter fit, together with their current experimental values \cite{Lenz:2010gu}.
As shown in the Table, $\left\vert V_{ud}\right\vert $,
$\left\vert V_{us}\right\vert $, $\left\vert V_{cd}\right\vert $, $\left\vert V_{cs}\right\vert $ and $\left\vert V_{tb}\right\vert $
are in excellent agreement with the experimental data, $\left\vert V_{ub}\right\vert $ and $\left\vert V_{td}\right\vert $ are around $40\%$ larger and $30\%$ smaller, respectively, than their experimental values, and finally $\left\vert V_{cb}\right\vert $ and $\left\vert V_{ts}\right\vert $ are one and two orders of magnitude lower, respectively. However, besides the specific values of the CKM elements, it is important to study their sensitivity to changes in the profile parameters, in order to verify whether their values arise naturally or accidentally. We actually found that the mixing of the first two generations, namely $V_{us}$ and $V_{cd}$, is highly sensitive to the profiles: for example, a change of $\sim 20\%$ in $d_{u_L}$ can cause a decrease in $V_{us}$ from $\sim 0.2$ to values below $10^{-2}$.
This strong sensitivity is due to the exponential dependence of the quark mass matrix elements on the profile localization parameters. This high sensitivity shows that the model, although able to fit the masses rather well, is not a good predictor for the mixing angles.

\begin{table}[tbh]
\begin{center}
\begin{tabular}{c|l|l}
\hline\hline
CKM matrix element & Obtained Value & Experimental Value \\ \hline
$\bigl|V_{ud}\bigr|$ & \quad $0.974$ & \quad $0.97426\pm 0.00030$ \\ \hline
$\bigl|V_{us}\bigr|$ & \quad $0.225$ & \quad $0.22545\pm 0.00095$ \\ \hline
$\bigl|V_{ub}\bigr|$ & \quad $0.00594$ & \quad $0.00356_{-0.00020}^{+0.00015}$ \\ \hline
$\bigl|V_{cd}\bigr|$ & \quad $0.225$ & \quad $0.22529\pm 0.00077$ \\ \hline
$\bigl|V_{cs}\bigr|$ & \quad $0.974$ & \quad $0.973416_{-0.00018}^{+0.00021}$ \\ \hline
$\bigl|V_{cb}\bigr|$ & \quad $0.0011$ & \quad $0.0493_{-0.00528}^{+0.00075}$ \\ \hline
$\bigl|V_{td}\bigr|$ & \quad $0.00602$ & \quad $0.00861_{-0.00037}^{+0.00021}$ \\ \hline
$\bigl|V_{ts}\bigr|$ & \quad $0.000293$ & \quad $0.040681_{-0.00138}^{+0.00043}$ \\ \hline
$\bigl|V_{tb}\bigr|$ & \quad $0.999$ & \quad $0.999135_{-0.000018}^{+0.000057}$ \\\hline\hline
\end{tabular}%
\end{center}
\caption{Obtained and experimental values of the magnitudes of the CKM
matrix elements.}
\label{table2}
\end{table}

\section{Conclusions}\label{sec:conclusions}
In this paper we have investigated the hierarchy of quark masses and mixings
in a model based on a 5-dimensional spacetime with constant curvature of
Randall-Sundrum type with two branes, where electroweak symmetry breaking
is triggered at the TeV brane by the condensation of a 4th generation. In this framework where the Standard Model fields and the 4th generation quarks are identified with the quark zero modes, their masses arise from the 4th generation $T$ and $B$ quark condensates. The condensates are originated from the strong couplings of the corresponding quarks with the first KK excitation of gluons, in addition to the effective interactions they may have from the 5-D bulk. The strong couplings between the 4th generation quarks and the first KK excitation of gluons arise from their high localization in a common region of the 5th dimension, namely near the TeV brane. The SM quark masses, which depend on the quark zero mode localization profiles, are generated by the bulk four-fermion interactions involving two SM quarks and two condensing 4th generation quarks.

In our first attempt, we freely explore the parameter space, and found that a moderate hierarchy among the four-fermion bulk interaction coefficients is strictly necessary to reproduce a realistic pattern for quark masses and mixings. This required hierarchy is clearly an indication that non trivial features of the underlying physics in 5-D are necessary.

In order to study these features in a more systematic way, we proceeded to impose restrictions to the bulk parameters by imposing a $U(1)$ symmetry, under which the fermion fields transform. This $U(1)$ symmetry thus creates textures in the quark mass matrices.  This framework generates non-symmetric modified Fritzsch-like textures for the ``up" and ``down" type SM quark matrices, and reproduces reasonably well the actual pattern of quark masses and mixings in the SM.
However, while the masses are reproduced quite naturally, the generated CKM mixing in the model
is very sensitive to the fine adjustment of the localization profiles. In this sense, the model is not a good predictor for the actual values of the CKM mixing matrix.

As a general conclusion, we found that, in a model that considers dynamical electroweak symmetry breaking triggered by quark condensation in warped 5-D spacetime, the actual values of the SM quark masses and mixings are not well reproduced by purely considering localization of the field profiles along the 5th dimension. We show it is necessary to consider details of the underlying physics in 5-D as well. This underlying physics is manifested as coefficients of effective four-fermion operators, and the details are exhibited as non-trivial patterns among these coefficients.

\section*{Acknowledgments}

This work is supported in part by Conicyt, Chile grant  ``Institute for advanced studies in Science and TechnologyÓ ACT-119.
 AZ also acknowledges support from Conicyt grant ``Southern Theoretical Physics Laboratory'' ACT-91 and Fondecyt grant 1110167.


\begin{thebibliography}{99}
\bibitem{PDG}K. Nakamura et al. (Particle Data Group), J. Phys. \textbf{G 37}, 075021 (2010)
\bibitem{Fritzsch} H. Fritzsch, {\ Phys. Lett.} \textbf{B 70}%
, 436 (1977); {\ Phys. Lett.} \textbf{B 73}, 317 (1978); {\ Nucl. Phys.}
\textbf{B 155}, 189 (1979); H.~Fritzsch and J.~Plankl, Phys.\ Lett.\ B
\textbf{237}, 451 (1990);  
D.~s.~Du and Z.~z.~Xing,
Phys.\ Rev.\ D \textbf{48}, 2349 (1993).
\bibitem{Fx} H.~Fritzsch and Z.~Xing, Phys.\ Lett.\ B
\textbf{555}, 63 (2003) [arXiv:hep-ph/0212195]; Prog.\ Part.\ Nucl.\ Phys.\
\textbf{45}, 1 (2000) [arXiv:hep-ph/9912358]; Nucl.\ Phys.\ B \textbf{556},
49 (1999) [arXiv:hep-ph/9904286]; Phys.\ Lett.\ B \textbf{353}, 114
(1995)[arXiv:hep-ph/9502297].
\bibitem{Xing2011}Z.~z.~Xing, D.~Yang, S.~Zhou, [arXiv:hep-ph/1004.4234v2];
A. C. B. Machado, J. C. Montero, V. Pleitez, [arXiv:hep-ph/1108.1767];
J.~E.~Kim, M.~S.~Seo, 	JHEP  \textbf{1102} (2011) 097, [arXiv:hep-ph/1005.4684].
\bibitem{Matsuda}H. ~Nishiura, K.~Matsuda, T.~Kikuchi and T.~Fukuyama,
Phys.\ Rev.\ D \textbf{65}, 097301 (2002) [arXiv:hep-ph/0202189];
K.~Matsuda and H.~Nishiura,
Phys.\ Rev.\ D \textbf{69}, 053005 (2004) [arXiv:hep-ph/0309272];
\bibitem{Branco}
  G.~C.~Branco, D.~Emmanuel-Costa and R.~Gonz\'alez Felipe,
  Phys.\ Lett.\  B {\bf 477}, 147 (2000)
  [arXiv:hep-ph/9911418];
  G.~C.~Branco, M.~N.~Rebelo and J.~I.~Silva-Marcos,
  Phys.\ Lett.\  B {\bf 597}, 155 (2004)
  [arXiv:hep-ph/0403016];
  G.~C.~Branco, L.~Lavoura and J.~P.~Silva,
  Int.\ Ser.\ Monogr.\ Phys.\  {\bf 103}, 1 (1999).
\bibitem{Zhou}
 Y.~F.~Zhou,
  J.\ Phys.\ G {\bf 30}, 783 (2004)
  [arXiv:hep-ph/0307240];
\bibitem{Carcamo}
 A.~E.~C\'arcamo, R.~Mart\'inez and J.~A.~Rodr\'iguez,
  Eur.\ Phys.\ J.\  C {\bf 50}, 935 (2007)
  [arXiv:hep-ph/0606190],
  AIP Conf.\ Proc.\  {\bf 1026} (2008) 272.
\bibitem{String} K.~S.~Babu and R.~N.~Mohapatra,
Phys.\ Rev.\ Lett.\ \textbf{74}, 2418 (1995) [arXiv:hep-ph/9410326].

\bibitem{GUT} R.~Barbieri, G.~R.~Dvali, A.~Strumia, Z.~Berezhiani and
L.~J.~Hall,
Nucl.\ Phys.\ B \textbf{432}, 49 (1994) [arXiv:hep-ph/9405428];
Z.~Berezhiani,
Phys.\ Lett.\ B \textbf{355}, 481 (1995) [arXiv:hep-ph/9503366];
H.~C.~Cheng,
Phys.\ Rev.\ D \textbf{60}, 075015 (1999) [arXiv:hep-ph/9904252];
A.~E.~C\'arcamo Hern\'andez and Rakibur Rahman [arXiv:hep-ph/1007.0447].

\bibitem{Extradim}Y.~Bai, M.~Carena and E.~Pont\'on,
Phys.\ Rev.\ \textbf{D81}, 065004 (2010) [arXiv:0809.1658 [hep-ph]];
 N.~Rius and V.~Sanz,
  Phys.\ Rev.\ D \textbf{64}, 075006 (2001) [arXiv:hep-ph/0103086]; B.~A.~Dobrescu, 
Phys.\ Lett.\ B \textbf{461}, 99 (1999) [arXiv:hep-ph/9812349];
K.~Agashe, R.~Contino and A.~Pomarol,
Nucl.\ Phys.\ B \textbf{719}, 165 (2005), [arXiv:hep-ph/0412089];
G.~Cacciapaglia, C.~Csaki, J.~Galloway, G.~Marandella, J.~Terning and A.~Weiler,
JHEP \textbf{0804}, 006 (2008), [arXiv:hep-ph/0709.1714];
H.~Ishimori, Y.~Shimizu, M.~Tanimoto and A.~Watanabe,
Phys.\ Rev.\ D \textbf{77}, 045027 (2008) [arXiv:0710.4344 [hep-th]].

\bibitem{Ncopies}A.~E.~C\'arcamo Hern\'andez, Sergey~Kovalenko and Ivan~Schmidt, to appear.

\bibitem{Chivukula} R. Sekhar Chivukula, [arXiv:hep-ph/00112641v1]; C.~T.~Hill and E.~H.~Simmons,
Phys.\ Rept.\ \textbf{381}, 235 (2003), [Erratum-ibid.\ \textbf{390}, 553
(2004)], [arXiv:hep-ph/0203079]; C.~T.~Hill, 
Phys.\ Lett.\ B \textbf{345}, 483 (1995), [arXiv:hep-ph/9411426]; C.~T.~Hill,
Phys.\ Lett.\ B \textbf{266}, 419 (1991).

\bibitem{Burdman1} Gustavo Burdman and Leandro Da Rold, JHEP \textbf{0712}
(2007) 086 [arXiv:hep-ph/0710.0623].

\bibitem{rs} L.~Randall and R.~Sundrum,
Phys.\ Rev.\ Lett.\ \textbf{83}, 3370 (1999), [arXiv:hep-ph/9905221].

\bibitem{Alok} Ashutosh Kumar Alok, Amol Dighe and David London, Phys.\
Rev.\ D \textbf{83}, 073008 (2011) [arXiv:hep-ph/1011.2634].

\bibitem{kribs} G.~D.~Kribs, T.~Plehn, M.~Spannowsky and T.~M.~P.~Tait, Phys.\
Rev.\ D \textbf{76}, 075016 (2007) [arXiv:hep-ph/0706.3718].

\bibitem{Erler} Jens Erler and Paul Langacker PRL \textbf{105}, 031801
(2010) [arXiv:hep-ph/1003.3211].

\bibitem{adms} K.~Agashe, A.~Delgado, M.~J.~May and R.~Sundrum,
JHEP \textbf{0308}, 050 (2003), [arXiv:hep-ph/0308036].

\bibitem{higgsless} C.~Csaki, C.~Grojean, L.~Pilo and J.~Terning,
Phys.\ Rev.\ Lett.\ \textbf{92}, 101802 (2004) [arXiv:hep-ph/0308038].

\bibitem{acdp} K.~Agashe, R.~Contino, L.~Da Rold and A.~Pomarol,
Phys.\ Lett.\ B \textbf{641}, 62 (2006), [arXiv:hep-ph/0605341].

\bibitem{delocal} G.~Cacciapaglia, C.~Csaki, C.~Grojean and J.~Terning,
Phys.\ Rev.\ D \textbf{71}, 035015 (2005) [arXiv:hep-ph/0409126].

\bibitem{Tasi}T.~Gherghetta,
[arXiv:hep-ph/1008.2570].

\bibitem{Burdman2} Gustavo Burdman and Carlos E. F. Haluch,
[arXiv:hep-ph/1109.3914v1];
 M.~A.~Luty,
Phys.\ Rev.\ D \textbf{41}, 2893 (1990).

\bibitem{bhl} W.~A.~Bardeen, C.~T.~Hill and M.~Lindner,
Phys.\ Rev.\ D \textbf{41}, 1647 (1990).

\bibitem{Gherghetta:2000qt}
  T.~Gherghetta, A.~Pomarol,
  Nucl.\ Phys.\  {\bf B586}, 141-162 (2000).
  [hep-ph/0003129].

\bibitem{Xing} Zhi-zhong Xing, He Zhang and Shun Zhou, Phys.\ Rev.\ D
\textbf{77}, 113016 (2008) [arXiv:hep-ph/0712.1419].

\bibitem{Lenz:2010gu} A.~Lenz, U.~Nierste, J.~Charles, S.~Descotes-Genon,
A.~Jantsch, C.~Kaufhold, H.~Lacker, S.~Monteil \textit{et al.},
Phys.\ Rev.\ \textbf{D83 }, 036004 (2011) [arXiv:1008.1593 [hep-ph]].

\end{thebibliography}
\end{document}